\documentstyle[12pt,epsf]{article}

\def\fun#1#2{\lower3.6pt\vbox{\baselineskip0pt\lineskip.9pt
        \ialign{$\mathsurround=0pt#1\hfill##\hfil$\crcr#2\crcr\sim\crcr}}}

\renewcommand\({\left(}
\renewcommand\){\right)}

\newcommand\eq[1]{Eq.~(\ref{#1})}
\newcommand\eqs[2]{Eqs.~(\ref{#1}) and (\ref{#2})}
\newcommand\eqss[3]{Eqs.~(\ref{#1}), (\ref{#2}) and (\ref{#3})}

\newcommand\pa{\partial}

\newcommand\ee{\end{equation}}
\newcommand\be{\begin{equation}}
\newcommand\eea{\end{eqnarray}}
\newcommand\bea{\begin{eqnarray}}

\newcommand\GeV{\,\mbox{GeV}}

\newcommand\mpl{M_{\rm P}}

\newcommand\lsim{\mathrel{\rlap{\lower4pt\hbox{\hskip1pt$\sim$}}
    \raise1pt\hbox{$<$}}}
\newcommand\gsim{\mathrel{\rlap{\lower4pt\hbox{\hskip1pt$\sim$}}
    \raise1pt\hbox{$>$}}}

\def\dslash{\not{\hbox{\kern-2pt $\partial$}}}
\def\Dslash{\not{\hbox{\kern-4pt $D$}}}
\def\Oslash{\not{\hbox{\kern-4pt $O$}}}
\def\Qslash{\not{\hbox{\kern-4pt $Q$}}}
\def\pslash{\not{\hbox{\kern-2.3pt $p$}}}
\def\kslash{\not{\hbox{\kern-2.3pt $k$}}}
\def\qslash{\not{\hbox{\kern-2.3pt $q$}}}
 \newtoks\slashfraction
 \slashfraction={.13}
 \def\slash#1{\setbox0\hbox{$ #1 $}
 \setbox0\hbox to \the\slashfraction\wd0{\hss \box0}/\box0 }
 
\def\ee{\end{equation}}
\def\be{\begin{equation}}

\def\calp{{\cal P}}
\def\calr{{\cal R}}

\newcommand\sub[1]{_{\rm #1}}

\newcommand\msinf{M\sub{inf}}

\begin{document}

\begin{flushright}
LANCS-TH/9818
\\hep-ph/9809310\\
(September 1998)
\end{flushright}
\begin{center}
{\Large \bf
Observational constraints on an inflation \\ model with a running mass}

\vspace{.3in}
{\large\bf  Laura Covi, David H.~Lyth and Leszek Roszkowski}

\vspace{.4 cm}
{\em Department of Physics,\\
Lancaster University,\\
Lancaster LA1 4YB.~~~U.~K.}

\vspace{.4cm}
{\tt E-mails: covi@virgo.lancs.ac.uk\ lyth@lavhep.lancs.ac.uk \
lr@virgo.lancs.ac.uk }
\end{center}

\vspace{.6cm}
\begin{abstract}
We explore a model of inflation where the inflaton mass-squared
is generated at a high scale 
by gravity-mediated soft supersymmetry breaking,
and runs at lower scales to the small value required for slow-roll 
inflation. The running is supposed to come from the coupling of the 
inflaton to a non-Abelian gauge field. In contrast with earlier work,
we do not constrain 
the magnitude of the supersymmetry breaking scale, 
and we find that the model might
work even if squark and slepton masses come from gauge-mediated supersymmetry 
breaking. 
With the inflaton and gaugino masses in the 
expected range, and $\alpha=g^2/4\pi$
in the range $10^{-2}$ to $10^{-3}$ (all at the high scale)
the model
can give the observed cosmic microwave anisotropy, and a spectral index 
in the observed range. The latter has significant variation with scale,
which can confirm or rule out the model in the forseeable future.

\end{abstract}
 
\section{Introduction}

It is  well known \cite{ll2,p97toni}
that slow-roll inflation requires
the flatness conditions $\epsilon\ll 1$ and
$|\eta|\ll 1$ on the potential, where
\bea
\epsilon &\equiv& \frac12 \mpl^2 \( \frac{V'}{V} \)^2 \label{eps} \\
\eta &\equiv& \mpl^2{V^{\prime\prime}\over V} \,, \label{eta}
\eea
and $M_{Pl} \equiv (8\pi G)^{-1/2} = 2.4\times 10^{18}\GeV$.
The first condition is automatically satisfied near a maximum
or minimum of the potential, where inflation is usually supposed
to take place, but the second is problematic.
A generic supergravity theory gives all scalar fields
\cite{dine,coughlan}, and in particular the inflaton
\cite{cllsw,ewansgrav} a mass-squared with magnitude 
at least of order $V/\mpl^2$, which would spoil this condition.
Yet
supergravity is generally considered as the appropriate framework for
a description of the fundamental interactions, and in particular
for the description of their scalar potential.

Very few proposals have been put forward to solve this problem which
would not be relying on some sort of fine tuning~\cite{p97toni}. Our aim is
to investigate in detail the proposal of 
Stewart~\cite{ewanloop1,ewanloop2} in
which loop corrections can flatten the inflaton
potential without any significant fine-tuning.
We follow him in assuming that the dominant corrections
come from a single gauge coupling of the inflaton, and explore the
region of parameter space which is allowed by the
observed magnitude and spectral index
of the curvature perturbation.

In the next section we give the general picture.
In Section 3 we
write down the Renormalization Group Equations (RGE's)
giving the scale-dependence of the inflaton mass, and hence 
its potential $V(\phi)$. In Section 4 we use the slow-roll approximation 
to derive an analytic expression for 
the number $N(\phi)$  of $e$-folds between a given epoch and 
the end of slow-roll inflation.
In Section 5 we derive the predictions of the model for the 
spectrum of the curvature perturbation, 
and for its dependence
on the comoving scale as specified by the spectral index $n(k)$.
They depend on five parameters; the inflaton mass $m_0$, the gaugino 
mass $\tilde m_0$ and the gauge coupling $\alpha_0$
(all evaluated at the Planck scale, as indicated by the subscript zero, 
with the last two multiplied by 
group-theoretic factors of order 1), the magnitude $V_0$ of the 
inflaton potential, and the number $N\lsim 55$
of $e$-folds of slow-roll inflation after a typical cosmological
scale leaves the horizon. 
Imposing the COBE measurement of the spectrum on large scales, 
and the 
observational constraint $|1-n|<0.2$ over the whole range of 
cosmological scales, we find an allowed region of parameter space that
includes the theoretically expected one.
In Section 5 we examine various consistency conditions on the 
calculation. They are generally satisfied in the 
allowed region calculated in Section 4.
Finally, we comment on the significance of the results and point out
that observation will confirm or exclude the model in the forseeable
future.

\section{The general picture}

We adopt the model of Stewart \cite{ewanloop2}
(see also the review \cite{p97toni}).
Slow-roll inflation occurs, with the
following  
Renormalization Group improved potential
for the canonically normalized inflaton field $\phi$;
\be
V =V_0 + \frac{1}{2} m^2 (\phi) \phi^2 + \cdots  \,.
\label{vinf}
\ee
The constant term $V_0$ is supposed to dominate at all relevant field 
values. Non-renormalizable terms, represented by the dots, 
give the potential a minimum at large $\phi$,
but they are supposed to be negligible 
during inflation.
The inflaton mass-squared $m^2(Q)$ depends on 
the renormalization scale $Q$, and following \cite{ewanloop1,ewanloop2}
we have taken
\be
Q=\phi
\ee
so that loop corrections will hopefully be small.\footnote
{\label{gfoot}
With a single coupling $g$ (as is the case for our model)
the one-loop correction vanishes 
at some value $Q\sim g\phi$. But we are interested in the regime $Q
\sim \mpl e^{-1/g}$, and as a result the fractional change
in observational quantities that would result from keeping the factor 
$g$ is of order $g^2$ (up to logs of
$g$). This is the same as the change that would come from including the 
two-loop correction, which justifies the simpler choice
$Q=\phi$.}

At the Planck scale $\phi=\mpl$,
$m^2(\phi)$ is supposed to be negative,\footnote
{\label{flatfoot}
Flat directions with negative mass-squared at a high scale have also
been studied using similar techniques in the context of colour and
electric charge breaking~\cite{fors}.
}
with the generic  magnitude
\be
|m^2_0 |=|m^2(M_P) | \sim V_0/\mpl^2
\label{mexpect1}
\ee
If there were no running, this would give $|\eta|\sim 1$,
preventing inflation. But at small field values, 
the RGE's drive $m^2(\phi)$ 
to small values, corresponding to $|\eta(\phi)|\ll 1$, and slow-roll
inflation can take place there. Within this region, there is a maximum 
of the potential, and we assume that the inflaton initially finds itself 
to the left of the maximum.\footnote
{The inflaton can arrive at the required value by tunneling from the 
minimum of $V$ located at large $\phi$. Such tunneling will certainly 
take place after a sufficiently long time, provided that $V$ is nonzero 
in the minimum. It is natural \cite{ewanloop2} to assume that this comes about
through the 
minimum not being the true vacuum, though even if it is the true vacuum
a cosmological constant is enough to cause tunneling \cite{gv}.}

Slow-roll inflation is assumed to continue until some epoch $\phi\sub{end}$,
when $\eta(\phi)$ becomes of order $+1$.
Then $\phi$ starts to 
oscillate around the origin, but inflation does not end because $V$ is 
still dominated by the constant term $V_0$.
In order to end inflation, we need a hybrid mechanism where
$V_0$ comes from the displacement of another field $\psi$
from its vacuum value. When the amplitude of the $\phi$
oscillation falls below some critical value $\phi\sub c$,
the other field $\psi$ rolls towards its vacuum value.

Inflation finally ends 
some number $N\sub{fast}$ of $e$-folds after the end of slow-roll
inflation, 
when $\phi$ and $\psi$ both settle down to their vacuum values.
For us this number is a free parameter, since we are not 
specifying the complete potential which anyhow would probably 
contain free parameters.  

Let us be more specific about the expected magnitude 
of $m^2$ at the Planck scale, \eq{mexpect1}.
We are assuming that the potential comes from the $F$ term of a supergravity 
theory, containing chiral superfields whose scalar components 
are complex fields $\phi_n$.
The real inflaton field $\phi$ is supposed to be a quasi-flat
direction in this multi-field space. 
Making the usual assumption that any fermion condensates can be
represented as a non-perturbative contribution to the superpotential,
the $F$ term has the familiar form
\bea
V &=& \msinf^4 
- 3e^{K/\mpl^2} \mpl^{-2} |W|^2 
\label{vexp}
\\
\msinf^4 &\equiv& 
 e^{K/\mpl^2} \left( W_n+\mpl^{-2}WK_n \right)
K^{n m^\star }
\left( W_m+\mpl^{-2}WK_m\right)^\star  
\,.
\label{vexp2}
\eea
Here $W$ is the superpotential, a holomorphic function of the
$\phi_n$, and $K$ is the K\"ahler potential, a real function of
the $\phi_n$ and their complex conjugates.
A subscript $n$ denotes $\pa/\pa \phi_n$ while $n^\star $
denotes $\pa/\pa \phi_n^\star $, and $K^{nm^\star }$ is the matrix inverse
of $K_{nm^\star }$.

The quantity $\msinf$ may be taken to define the scale of supersymmetry
breaking during inflation.
 (The similarly-defined quantity in the vacuum
is generally denoted by $M\sub S$.)
An essential assumption is
that the inflaton mass is negligible in the 
limit of global supersymmetry, and 
comes only from supergravity corrections.
The mass-squared of the inflaton (and of any other scalar field receiving its 
mass only from supergravity)
is then given by
\be
\mpl^2 m^2 = V \pm c\msinf^4 \,..
\label{mofms}
\ee
The first term comes \cite{cllsw,ewansgrav} from the factor $e^{K/\mpl^2}$.
The factor $c$ in the second term receives contributions from 
non-renormalizable terms in $K$ \cite{p97toni}. 
In units of $\mpl$, the coefficients of these terms are supposed to
be roughly of order 1 at the Planck scale. This gives
$c\sim 1$,
within some uncertainty factor $10^\Delta$ that is a matter 
of taste. 

We now make two crucial assumptions. One is that, in contrast with the case 
for the vacuum, there is no strong cancellation 
between the two terms of \eq{vexp}, so that 
$V\sim \msinf^4$. The validity of this assumption could be checked in a 
particular model. Second, there is no
strong cancellation between
the two terms of \eq{mofms}. This second assumption is a statement about
the non-renormalizable terms of $K$, whose status is hard to judge.
A strong cancellation might happen by accident, or because we are 
dealing with a model in which $K$ and $W$ 
have special forms that ensure the cancellation
\cite{p97berk,p97toni}.\footnote
{This class of models does not include those with a superpotential
that is linear (during inflation) in the inflaton field
\cite{linear}. 
In these models there is indeed an identifiable contribution $-V$ in 
the second term of \eq{mofms} \cite{cllsw}, but
for every field with a value of order $\mpl$ there is another 
contribution of order $+V$, coming from the term $e^K K_n K_m^\star  K^{nm^\star } 
\mpl^{-4}|W|^2$ in \eq{vexp2}.}
For the purpose of this paper, we are taking the view that a strong 
cancellation is unlikely.

With these assumptions, 
the expected
value of the inflaton mass-squared at the Planck scale is 
\be
 1 \lsim \mpl^2|m^2(\mpl)|/V_0 \lsim 10^\Delta
\,,
\label{mexpect}
\ee
with the value $\sim 1$ regarded as the most likely,
and values $\ll 1$ regarded as {\em un}likely.

It is instructive to compare this analysis with the usual one,
which applies in the vacuum. Here, the mass-squared  of a scalar
field, if it comes only from supergravity, is 
given by the analogue of \eq{mofms}, $\mpl^2 m^2= \pm cM\sub S^4$,
the supersymmetry breaking scale $M\sub S$ being related to the potential
by \eq{vexp} with $\msinf$ replaced by $M\sub S$.
The  contribution to $\mpl^2 m^2$ from $V$ is now absent, because
$V$ vanishes in the vacuum. As a result it is equally reasonable to 
expect $m^2$ to be a factor $10^\Delta$ above {\em or below} 
the estimate $\mpl^2 m^2 \sim  M\sub S^4$. There is no prejudice against
values below, as is the case during inflation.

In summary, we have seen that achieving
the slow-roll requirement $\mpl^2|m^2|\ll V_0$
on the inflaton mass
is {\em not} just a matter of tuning something a bit below
its expected value. Rather, one has to invoke a strong
cancellation between different contributions.
The idea of the present model is to avoid that, by running the inflaton 
mass.

In the limit where the inflaton mass has negligible scale dependence,
we arrive at the tree-level model of Randall, Soljacic and Guth
\cite{lisa}, which invoked an accidental cancellation to keep the 
inflaton mass sufficiently small.
Apart from avoiding the cancellation, the strong running of the inflaton 
mass invoked in the present model has another
advantage \cite{ewanloop2}, which is that slow-roll
inflation can (and is supposed to) end before $\phi$ reaches the critical value
$\phi\sub c$. This makes the predictions practically independent of
$\phi\sub c$. It also reduces, or maybe eliminates,
the dangerous spike in the spectrum
of the curvature perturbation, that is
generated in the tree-level model after slow-roll inflation ends.

The tree-level model \cite{lisa}, and Stewart's paper \cite{ewanloop2}, 
actually 
made two assumptions, beyond the ones mentioned already.
First, that the scales of supersymmetry breaking during
inflation and in the vacuum are similar, $M\sub S\sim 
\msinf\simeq V^{1/4}$. 
Second, that the squarks and sleptons of
the supersymmetric Standard Model also get their masses only from
supergravity (gravity-mediated supersymmetry breaking).
With these assumptions one will have
roughly $V_0^{1/4}\sim \sqrt{m\sub s\mpl}\sim 10^{10}\GeV$
where $m\sub s$ 
roughly of order $100\GeV$ is the expected  mass of squarks and sleptons.
We shall not make these assumptions, and leave
$V_0$ as an essentially free parameter. 
If we retain the first assumption, but assume \cite{p97toni}
that the squark and 
slepton masses of the 
supersymmetric Standard Model are 
gauge-mediated, 
we will have $10^5\GeV
\lsim V_0^{1/4}\lsim 10^{10}\GeV$. 
We stress that the latter case can only be realised when the inflaton sector is
different, and decoupled from, the visible sector, due to our
assumption that the inflaton mass is generated by gravity-mediated
soft supersymmetry breaking.

\section{The Renormalization Group Equations}

We assume that in the inflaton sector, the RGE's are those
of softly broken global supersymmetry.
To obtain the functional form of $m^2(Q)$ we follow 
Stewart \cite{ewanloop2} in assuming that the 
inflaton has a gauge coupling, which dominates the RGE's.
(This is a hybrid inflation
model, so the inflaton certainly has Yukawa couplings but they are 
assumed to have a negligible effect.) 

The RGE's will involve the gaugino
mass $\tilde m(Q)$.
In contrast with the case of scalar masses, gaugino masses might
a priori be very small, because they
depend on the gauge kinetic function $f$.\footnote
{They vanish
if it is independent of the fields that break supersymmetry
(ie., have $W_n+\mpl^{-2}K_nW \neq 0$). This happens, for instance, 
with gaugino condensation
in weakly coupled string theory where $f$ is proportional
to the dilaton field and the latter does not break supersymmetry.}
Allowing for this possibility, the
expected value of the
gaugino mass at the Planck scale is therefore
$0\lsim (\mpl^2\tilde m^2(\mpl)/V_0)
\lsim 1$, or allowing some uncertainty
\be
0\lsim \frac{\mpl^2\tilde m^2(\mpl)}{V_0} \lsim 10^\Delta
 \,.
\label{mtilexpect}
\ee

We also need the gauge coupling at the Planck scale, 
and we make the usual assumption that 
it is in the perturbative regime $\alpha(\mpl)\lsim 0.1$. On the other hand
one does not expect it to be many orders of magnitude smaller, 
under the usual supposition that gauge couplings at the high scale come 
more or less directly from something like string theory.
For definiteness, let us say that we expect
\be
10^{-3}\lsim \alpha(\mpl) \lsim 10^{-1}\,.
\label{alphaexpect}
\ee

{}From now on, we set $\mpl=1$ except where stated.
We work at one loop order,
and consider scales well above the relevant masses,
\be
\tilde m(Q), \ m(Q) \ll Q \label{tilmmreq} \,.
\ee
The RGE's have the same
form as the well-known
ones that describe the running of the squark masses with only QCD 
included, 
\bea
{d \alpha \over dt} &=& {b \over 2 \pi } \alpha^2 
\label{RGE-alpha}
\\
{d \over dt} \left( {\tilde m \over \alpha} \right) &=& 0 
\label{RGE-gaugino}
\\
{d m^2_{\phi} \over dt} &=& - {2 c \over \pi} \alpha \tilde m^2 
\label{RGE-inflaton}
\eea
Here 
$\alpha $ is related to the gauge coupling by
$\alpha = g^2/4\pi$, and 
$t \equiv \ln Q < 0$. The numbers $b$ and $c$ depend on the group;
$c$ is the Casimir 
quadratic invariant of the inflaton representation under the gauge group,
for instance 
$c = (N^2-1)/2N $ for any fundamental representation of $SU(N)$,
and 
$b = - 3 N + N\sub{f} $ 
for a supersymmetric $SU(N)$ with $N\sub{f}$ pairs of fermions
in the fundamental/antifundamental representation.
It is not clear whether one can construct a model in which the gauge 
group is actually the $SU(3)$ of QCD. 
All we assume for the moment is that $c$ and $-b$ are positive.

Using a subscript 0 to denote the
Planck scale $t=\ln\mpl=0$,
the solutions of equations~(\ref{RGE-alpha})-(\ref{RGE-inflaton}) are 
\bea
\alpha (t) &=& {\alpha_0 \over 1  - {b \alpha_0 \over 2 \pi} t}
\label{sol-alpha}
\\
\tilde m (t) &=& {\tilde m_0 \over 1- {b \alpha_0 \over 2 \pi} t} 
\label{sol-gaugino}
\\
m^2(t) &=& m^2_{0} + {2 c\over b}\tilde m^2_0 
{ \left[ 1- {1 \over \left[ 1-{b\alpha_0\over 2\pi}
t\right]^2}\right] }
\label{sol-inflaton}
\eea
We are taking $b<0$ and $c>0$, which means that $\alpha$,
$\tilde m$ and $m$ all increase as the scale decreases, and that
$m^2$ can cross from negative to positive values.

\subsection{The field dynamics}

We are making the identification $\phi=Q$, corresponding to 
$\ln\phi=t$.
It is useful to define the following quantities
\bea
\mu^2 &\equiv& -m^2/V_0 \\
A &=&  {2 c\over |b|} {\tilde m^2\over V_0} 
\label{param} \\
\tilde \alpha &=& {|b|\alpha \over 2\pi } 
\,.
\eea
(Remember that we are setting $\mpl=1$.)
For $SU(N)$ (and $b<0$ as we are assuming) one has
\bea
|b| &=& 3N-N\sub f \\
\frac{2c}{|b|} &=& \frac{1-N^{-2}}{3-N\sub f/N}  \,.
\eea
At least in this case, one can expect $A\simeq \tilde m^2/V_0$
and $\tilde\alpha\simeq \alpha$. 
With this assumption, 
the expected values at 
$\mpl$ corresponding to 
\eqss{mexpect}{mtilexpect}{alphaexpect} are
\bea
1 &\lsim& \mu_0^2 \lsim 10^\Delta \label{muexpect} \\
 0 &\lsim& A_0 \lsim 10^\Delta \label{aexpect} \\
10^{-3} &\lsim& \tilde \alpha_0 \lsim 10^{-1} \,..
\eea
Choosing $\mu^2_0 = A_0=1$ and $\tilde\alpha_0 = 0.01$,
we plot the potential in Figure 1.

Defining
\be
\gamma \equiv  \mu^2 + {1\over 2} {d\mu^2\over dt} 
\label{gamma} \,,
\ee
the derivatives of the potential \eq{vinf} with respect to $\phi$ are
\bea
V^{\prime} &=& - V_0\phi \left[\mu^2 + {1\over 2} {d\mu^2\over dt} \right] = 
V_0 \phi\ \gamma
\label{V-prime}\\
V^{\prime\prime} &=& - V_0  \left[\mu^2 + {3\over 2} {d\mu^2\over dt} +
{1\over 2} {d^2\mu^2\over dt^2}
\right]  \,.
\label{eps-eta}
\eea
Each derivative of 
$\mu^2$ is suppressed by a factor $\tilde \alpha_0$ with respect to $\mu^2$.
As a result \cite{ewanloop1} 
the flatness conditions \eqs{eps}{eta} are satisfied
in a region around $V'=0$, allowing inflation to take place there.

\begin{figure}
\begin{center}
\epsfysize= 2.5in %2.75in
%\epsffile{pot.eps}
\centerline{\epsfbox{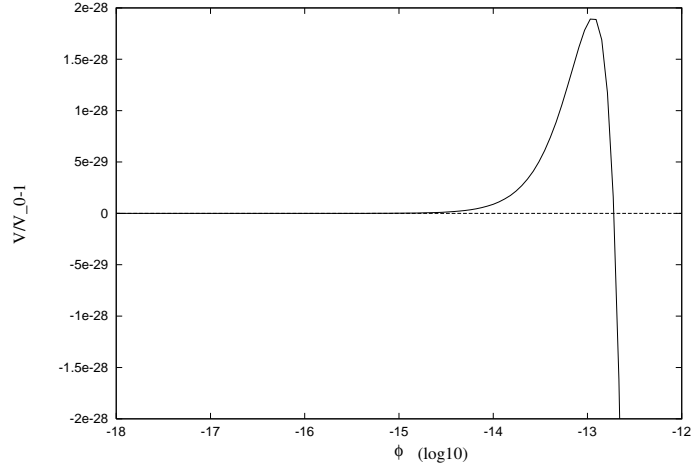}}
\caption{One loop potential $V/V_0-1$ for $\mu^2_0 =A_0
= 1$ and $\tilde\alpha_0 = 0.01$,
as a function of the scale $Q=\phi$.}
\label{fig-potential}
\end{center}
\end{figure}

\begin{figure}
\begin{center}
\epsfysize= 2.5in %2.75in
\centerline{\epsfbox{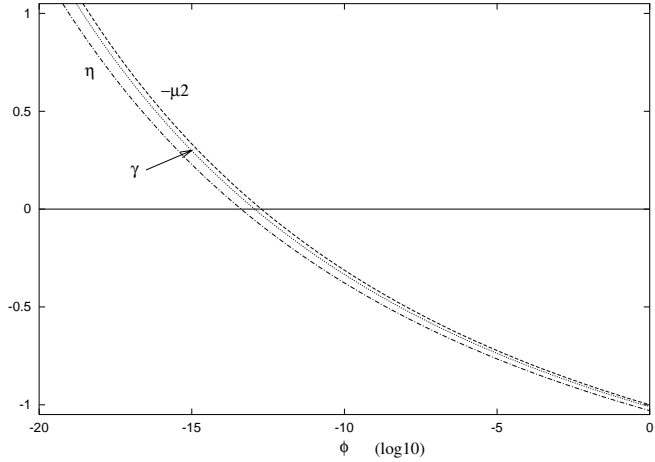}}
\caption{The functions $\mu^2$, $\gamma$ and $\eta$
for $\mu^2_0 =A_0
= 1$ and $\tilde\alpha_0 = 0.01$,
as a function of the scale $Q=\phi$.}
\label{fig-mass}
\end{center}
\end{figure}

The flatness parameters are
\bea
\eta &\simeq& \gamma + {d\gamma\over dt} = 
- \left[\mu^2 + {3\over 2} {d\mu^2\over dt} + 
{1\over 2} {d^2\mu^2\over dt^2} \right] \label{etaexp} \\
\epsilon &=& \frac12 \gamma^2\phi^2 \ll |\eta| \,.
\eea
The last two terms of \eq{etaexp}
are generically small with respect to $\mu^2$, but not
in the region of interest $\mu^2 \simeq 0$.

In Figure 2 we plot the quantities $\mu^2$, $\gamma$ and $\eta$
for a representative case.

The classical equation of motion for the inflaton field is 
\be
\ddot \phi + 3 H \dot \phi + V^{\prime} = 0 \,,
\label{full}
\ee
where $H^2 = (V + \dot \phi^2/2)/3$ is the Hubble parameter.
For a wide range of initial conditions, the flatness conditions 
lead to the slow-roll approximation
\be
3 H \dot\phi = -V^\prime \,.
\ee
We assume that slow-roll inflation ends at $\phi\sub {end}$, 
defined by
\be
\eta(\phi\sub{end}) = 1. 
\ee

To impose observational constraints on the model, we shall need the 
number of $e$-folds $N(\phi)$ of slow-roll inflation, taking place
after a given epoch $\phi$. 
Using the slow-roll approximation and $t=\ln\phi$,
it is given by
\be
N = \int_{t\sub{end}}^{t} {dt \over \gamma} \ \ .
\label{N-sr}
\ee

To evaluate the integral, we define
\be
y\equiv (1+\tilde \alpha_0 t)^{-1} = 
(1+\tilde \alpha_0 \ln\phi)^{-1} \,.
\ee
As $\phi$ decreases from $\phi=\mpl=1$, $y$ increases from $y_0=1$.
We define the 
epoch $y_\star $ 
by $\gamma  =0(=V')$, and the epoch $y_{\star \star }$ by
$\mu^2 =0(=m^2) $.
The latter is given by
\be
y_{\star\star} = \sqrt{1+{\mu_0^2\over A_0}}  \,.
\ee

In terms of these quantities
\bea
\tilde \alpha(y) &=& \tilde \alpha_0 y \\
A (y) &=& A_0 y^2 \\
\mu^2 (y) &=& A_0 \left[ y_{\star\star}^2 -y^2\right] \\
\gamma (y) &=&  - A_0 \left[y_{\star\star}^2 - y^2 
+ \tilde\alpha_0 y^3 \right] \label{gamma-y} \\
\eta (y) &=&  - A_0 \left[ y_{\star\star}^2 - y^2 
+ 3 \tilde\alpha_0 y^3  - 3 \tilde\alpha_0^2 y^4 \right].
\eea 
Notice that $A$ and $-\mu^2$ differ only by a constant.

To first order in $\tilde \alpha_0$, the equation $\gamma=0$ has the 
roots $y_\star $ and $y_\pm$ given by
\bea
y_{\star} &\simeq&  y_{\star\star} 
(1+ {1\over 2} \tilde\alpha_0 y_{\star\star}) \,, 
\label{ystar}\\
y_\pm &=& {1 - \tilde\alpha_0 y_\star \over 2\tilde\alpha_0} 
\pm \sqrt{
{(1 + \tilde\alpha_0 y_\star)^2\over 4\tilde\alpha_0^2} 
- y_\star^2} \ \simeq \left\{ 
\begin{array}{c}
1/\tilde\alpha_0 \\ - y_\star
\end{array}
\right.
\label{roots}
\eea
We are assuming that the potential has a maximum, corresponding to
real $y_\star $. This implies that $y_+$ is also real
(because there is always a real root $y_-<0$).
The root $y_+$ represents a minimum of the potential, but since
it lies in the regime of strong coupling it can hardly 
be trusted. Whether or not it is a real effect, we need to assume
that slow-roll inflation ends before it is reached, 
corresponding to $y_+ > y\sub{end}$.\footnote
{One actually hopes that the minimum represented by $y_+$
is not a real effect, because otherwise
one needs to fine-tune the critical value $\phi\sub c$ to be near
the minimum (given our assumption 
$\phi\sub c<\phi\sub{end}$).}
According to \eq{ystar},
the maximum of the potential is close to the point where
$m^2$ passes through zero.

Using these expressions, 
\bea
N (y) &=&  - {1\over \tilde\alpha_0} \int_{y\sub{end}}^{y} 
{dy \over y^2\gamma(y)} \\
&=& {1\over A_0 \tilde\alpha_0} 
\left[ 
{1\over y_{\star\star}} \left( {1\over y } - 
{1\over y\sub{end}} \right)  + 
{1\over \tilde\alpha_0 y^2_+ (y_+ - y_-)(y_+ - y_\star)} 
\ln\left( {y_+ - y \over y_+ - y\sub{end}} \right) \right. + \nonumber\\
 &+& {1\over\tilde\alpha_0 y^2_- (y_+ - y_-)(y_\star - y_-)} 
\ln\left( {y - y_- \over y\sub{end} - y_-} \right) - \nonumber\\
&-& \left. {1\over \tilde\alpha_0 y^2_\star (y_+ - y_\star)(y_\star - y_-)} 
\ln\left( {y - y_\star \over y\sub{end} - y_\star} \right) 
\right]
\label{N-y}
\eea
Apart for the mild dependence hidden in $y\sub{end}$, the number of e-folds 
given by eq.~(\ref{N-y}) is inversely proportional to $A_0$; the dependence on 
$\tilde\alpha_0$ is more complex, but anyway a small coupling increases $N$.

We have solved the second order equation (\ref{full}) numerically in the
region of interest and compared the number of e-folds with the 
slow-roll result, \eq{N-y}.
The latter is a good 
approximation over all the range considered, 
giving at most an error 
$\Delta N=2$ to 3, and
we use it in what follows.
(In \cite{ewanloop1,ewanloop2} the slow-roll approximation was replaced 
by  a different approximation.)

\section{The observational constraints}

The vacuum fluctuation of the inflaton fields generates 
a primordial adiabatic density perturbation, whose spectrum 
$\calp_\calr$ is
given by
\be
\frac4{25}\calp_\calr\equiv
\delta_H^2(k) = \frac1{75\pi^2}\frac{V^3}{V'^2}
=
\frac1{150\pi^2}\frac{V}{\epsilon}.
\label{delh}
\ee
The right hand side is evaluated at the epoch of horizon exit,
$k=aH$. 

The spectrum is 
constrained by observations of galaxies and of the cosmic microwave
background (cmb) anisotropy \cite{ll2,p97toni}. 
These measurements explore scales (the `cosmological
scales') ranging from the size of the 
observable Universe down to around four orders of magnitude smaller.
Near the top end of the range the COBE measurement of the cmb
gives an accurate value \cite{bunn96} for the spectrum, $\delta_H
=1.91\times 10^{-5}$ (the uncertainty is of order $10\%$,
ignoring gravitational waves). 
The dependence of the spectrum on the comoving wavenumber $k$
is defined through the spectral index $n$;
\be
n(k) -1 \equiv
d\calp_\calr/d\ln k \,.
\ee
Assuming that the $n$ is slowly varying over the 
cosmological range, 
observation requires something like 
$|n-1|<0.2$.

The perturbation on a comoving scale with wavenumber $k/a(t)$
is generated at the epoch of horizon exit $k=aH$, which is proportional
to $a\propto e^{Ht}$ in the slow-roll approximation. Hence the 
four-decade span of 
cosmological scales corresponds to $\Delta N\simeq 
\ln 10^4\simeq 9$.
If there is no further inflation, the value of $N$ at the top end of the 
range is
\cite{ll2,p97toni}
\be
N\sub{COBE}=48-\ln(10^{10}\GeV/V_0^{1/4})
-\frac13 \ln( V_0^{1/4}/\rho_{\rm reh}^{1/4}) \,,
\ee
where $\rho_{\rm reh}$ is the energy density at the
`reheat' temperature marking the beginning of radiation 
domination. As the notation indicates, we take the top end of the range 
to be the scale probed by COBE, though the latter is actually somewhat 
lower. 

In the present model 
some number $N\sub{fast}$ of $e$-folds of inflation take place after 
slow-roll ends \cite{ewanloop2}. 
Also, there could 
one or more bouts of thermal inflation before final reheating
\cite{thermal}, each bout lasting for of order 10 $e$-folds.
These reduce $N\sub{cosm}$ by an amount $N\sub{fast}+N\sub{thermal}$,\footnote
{Actually, $N\sub{fast}$ is 
only a rough estimate of the relevant reduction in $N\sub{cosm}$,
because it holds only if the variation of $H$ during fast-roll inflation 
is ignored.}
so $N\sub{COBE}$ is a free parameter of the model subject to the constraint
\be
10\lsim N\sub{COBE}< 48-\ln(10^{10}\GeV/V_0^{1/4})
-\frac13\ln(V_0^{1/4}/\rho_{\rm reh}^{1/4}) \,.
\ee
(The lower limit comes from the requirement that the perturbation is 
generated on all cosmological scales.)

The COBE normalization corresponds to 
\be
{V^{3/2} \over V^{\prime} } \simeq 6 \times 10^{-4} \,,
\label{cobe}
\ee
and the spectral index is given by
\be
n -1= 2\eta\,.
\ee

With our potential, the COBE normalization gives
\be
{ V_0^{1/2} \over \gamma (y) } \exp \left[ { 1\over \tilde\alpha_0}
\left( 1 - {1\over y} \right) \right] = 6 \times 10^{-4} \,,
\label{cobe-y}
\ee
and 
\be
n-1 = -2A_0 \(y^2_{\star\star} - y^2 + 3 \tilde\alpha_0 y^3  - 
3 \tilde\alpha_0^2 y^4 \) \,.
\label{nofy}
\ee 

In these expressions we regard $y$ as a function of $N$
(the inverse of \eq{N-y}), and it should be evaluated at the 
epoch of horizon exit for the relevant scale, $k=aH$.
In \eq{cobe-y}, the value of $y$ is fixed, corresponding to
the COBE scale which is in the upper part of the cosmological range.

It is not possible to explore the parameter space analytically without 
simplifying assumptions; we will therefore solve the problem numerically
and explore the region in the theoretically allowed parameter space where 
the spectral index satisfies the experimental bound.

Taking $N\sub{COBE}=45$, we
show in Figure 3 some lines of constant $V_0$, in the $\mu_0^2$-$A_0$
plane for some choices of $\alpha_0$. We also show some lines 
of constant $n$, evaluated also at the COBE scale.

Keeping the choice $N\sub{COBE}=45$,
we show in Figure 4 the dependence of the spectral index
on the comoving scale, for fixed values of $A_0$
and $\mu_0$ and the same values of $\alpha_0$.
The variable used to specify the comoving wavenumber $k$ is 
$N(\phi)=\ln(k\sub{end}/k)$, where $\phi$ is the epoch of horizon exit
$k=aH$ and $k\sub{end}$ corresponds to the end of slow-roll inflation.
Cosmological scales correspond to $35\lsim N(\phi)\lsim 45$.

Quite generally, there is significant variation over the range of 
cosmological scales. For instance, one can show that at $n=1$,
\be
\frac{dn}{dN} \simeq - 8 A_0^2 \tilde\alpha_0^2 \( 1 + \frac{\mu_0^2}
{A_0} \)^3 \,.
\ee
Over the cosmological range $\delta N\simeq 10$ one typically
has at $\delta n\sim 0.1$, which should be detectable 
with the advent of the MAP satellite and new galaxy surveys.

One can use Figure 4 to determine the effect of 
changing $N\sub{COBE}$, on the curves of constant $n$ in
Figure 3. For example, each of them has $n(45)\simeq 1.0$,
and $n(30)\simeq 1.2$; therefore 
changing to $N\sub{COBE}\simeq 30$ corresponds to shifting
the curves $n=1.0$ in Figure
3 to the location of the curves $n=1.2$, and shifting
the curves $n=0.8$ to the location $n=1.0$.

\begin{figure}
\epsfysize= 2.2in
\centerline{\epsfbox{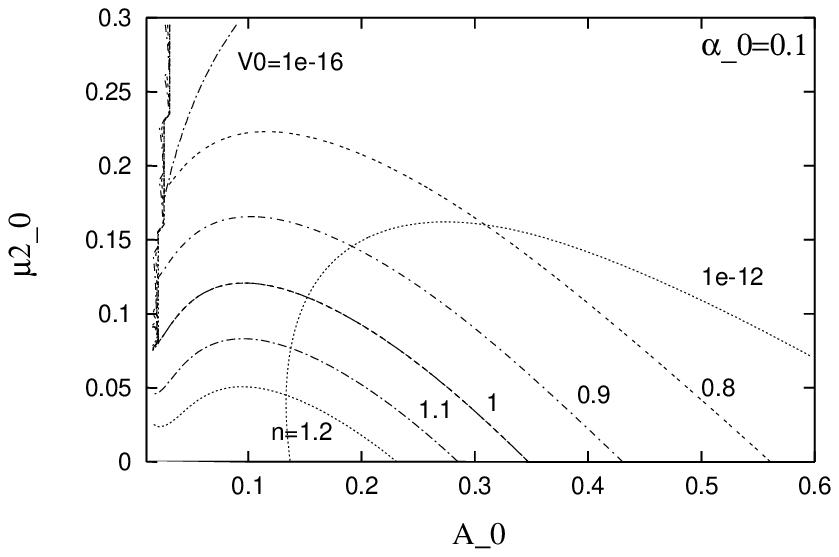}\epsfysize= 2.2in\epsfbox{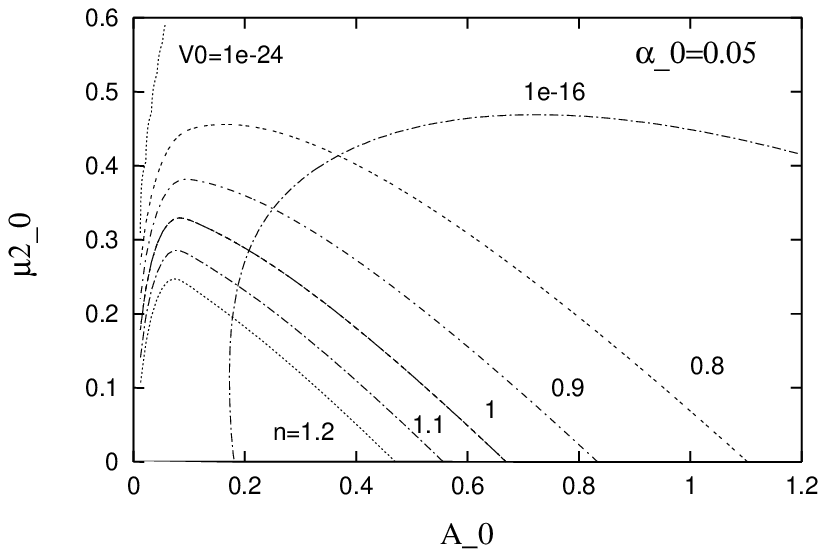}}
\epsfysize= 2.2in
\centerline{\epsfbox{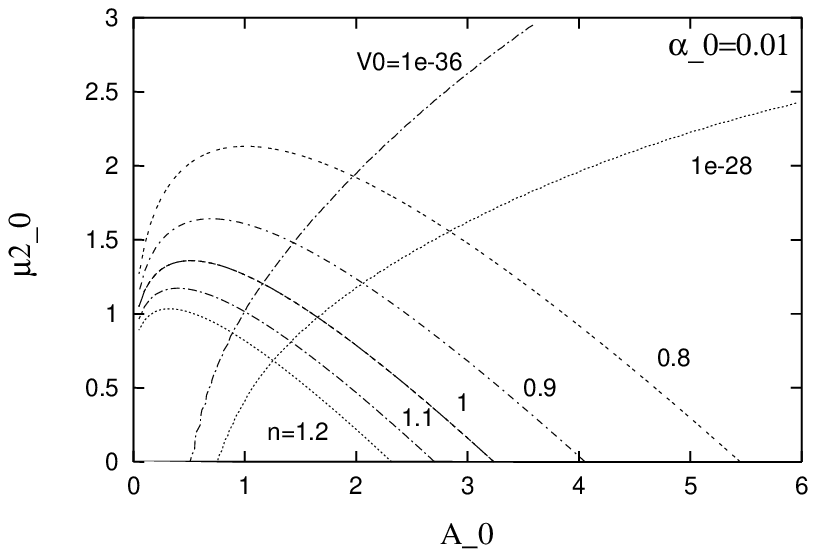}\epsfysize= 2.2in\epsfbox{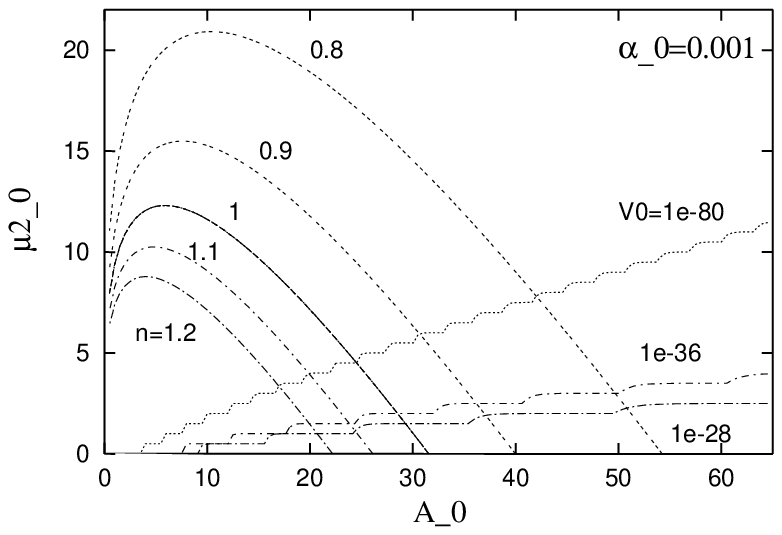}}
\caption{Contour levels of the spectral index and of $V_0$ in the 
$\mu^2_0-A$ plane for the different values of the coupling 
($\tilde\alpha_0 = 0.1,0.05,0.01,0.001$). Notice that for  
$\tilde\alpha_0 = 0.1$ (upper left plot) a small strip of the 
parameter space is excluded by the consistency constraints (see Fig.~5): 
the unlabelled vertical line on the left indicates such constraint 
and is not a contour level. In the other cases the whole parameter region 
displayed here is contained in the allowed region of Fig.~5.}
\end{figure}

\begin{figure}
\epsfysize= 3.75in
\centerline{\epsfbox{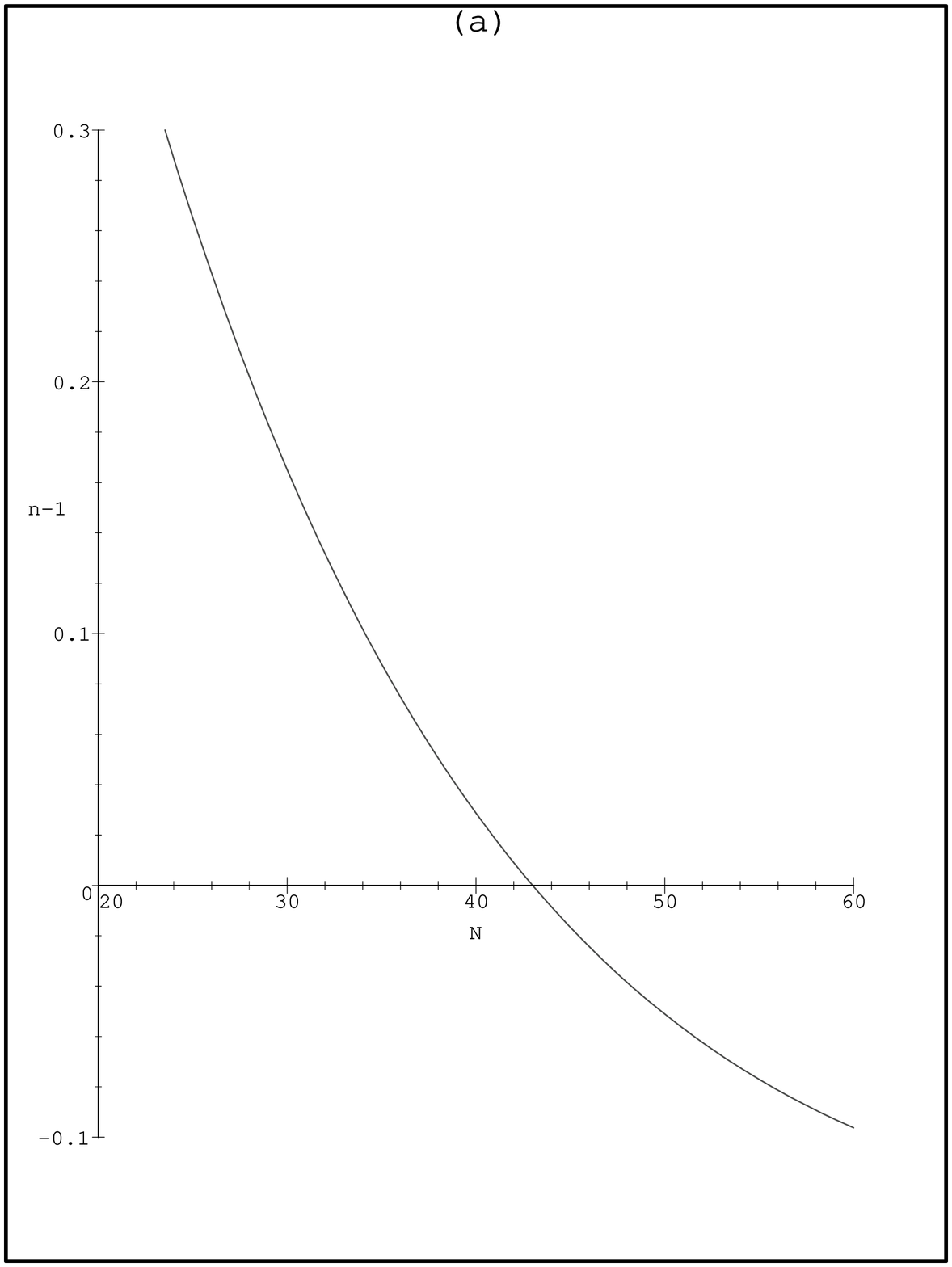}\epsfysize= 3.75in\epsfbox{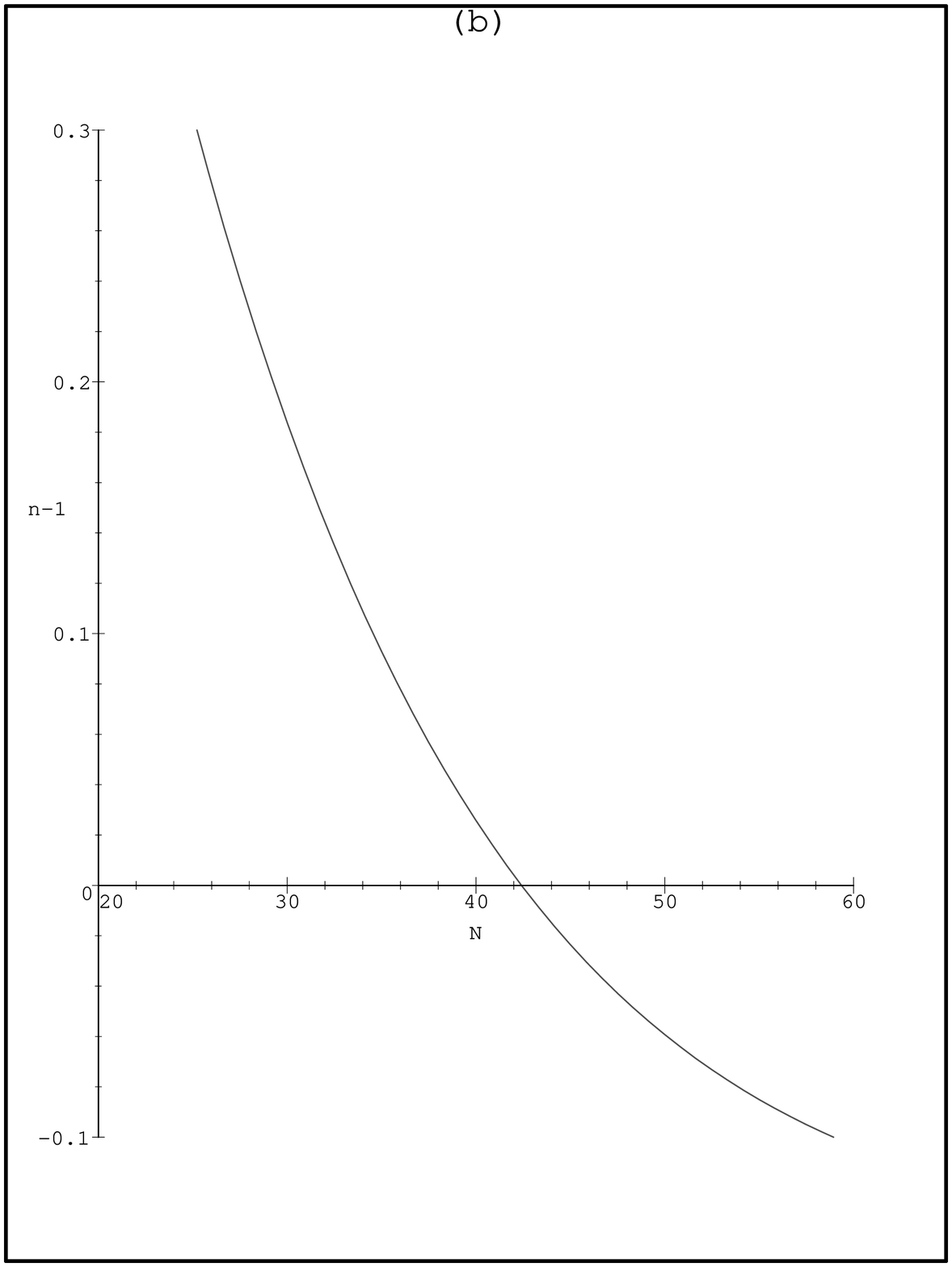}}
\epsfysize= 3.75in
\centerline{\epsfbox{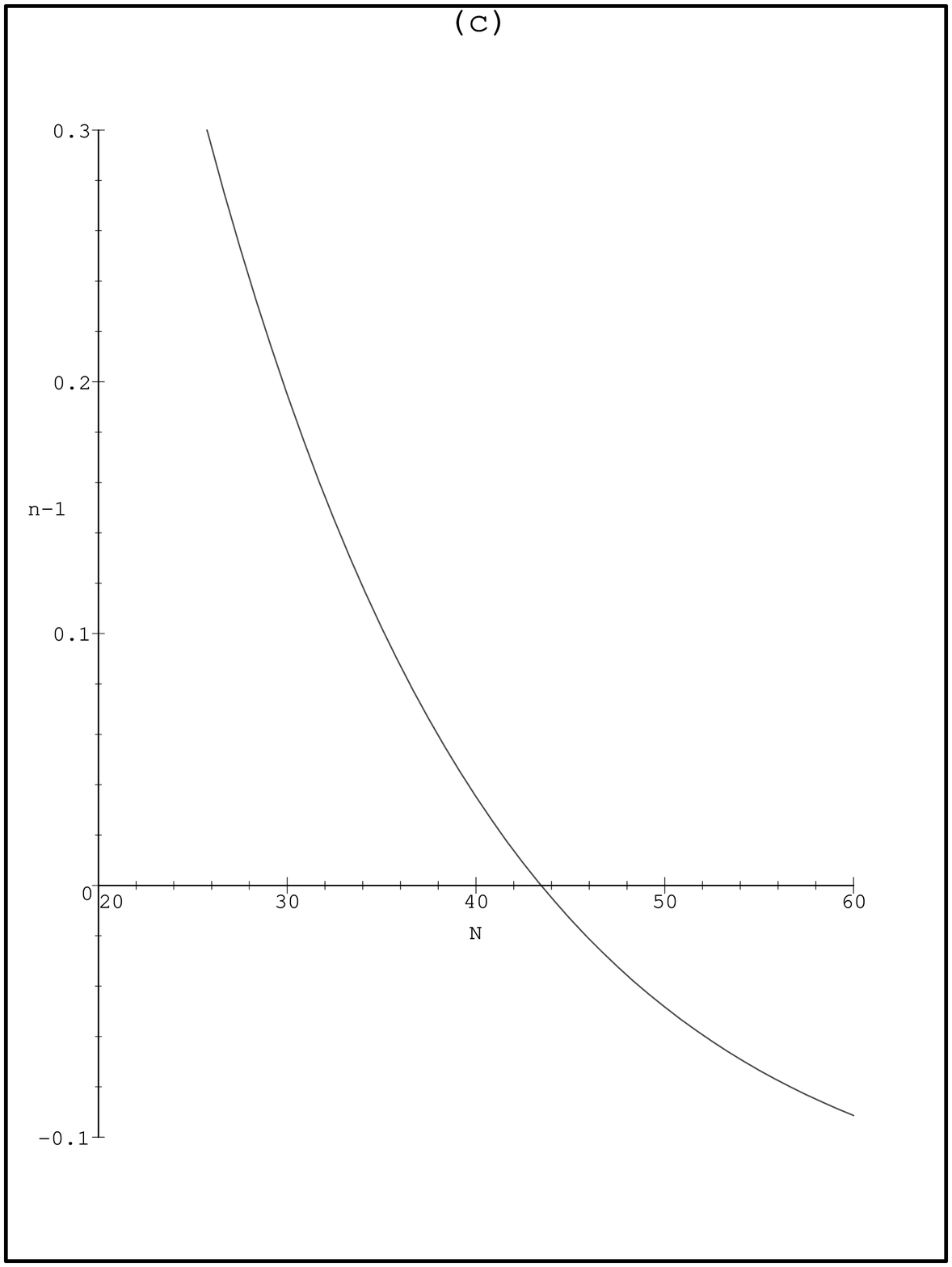}\epsfysize= 3.75in\epsfbox{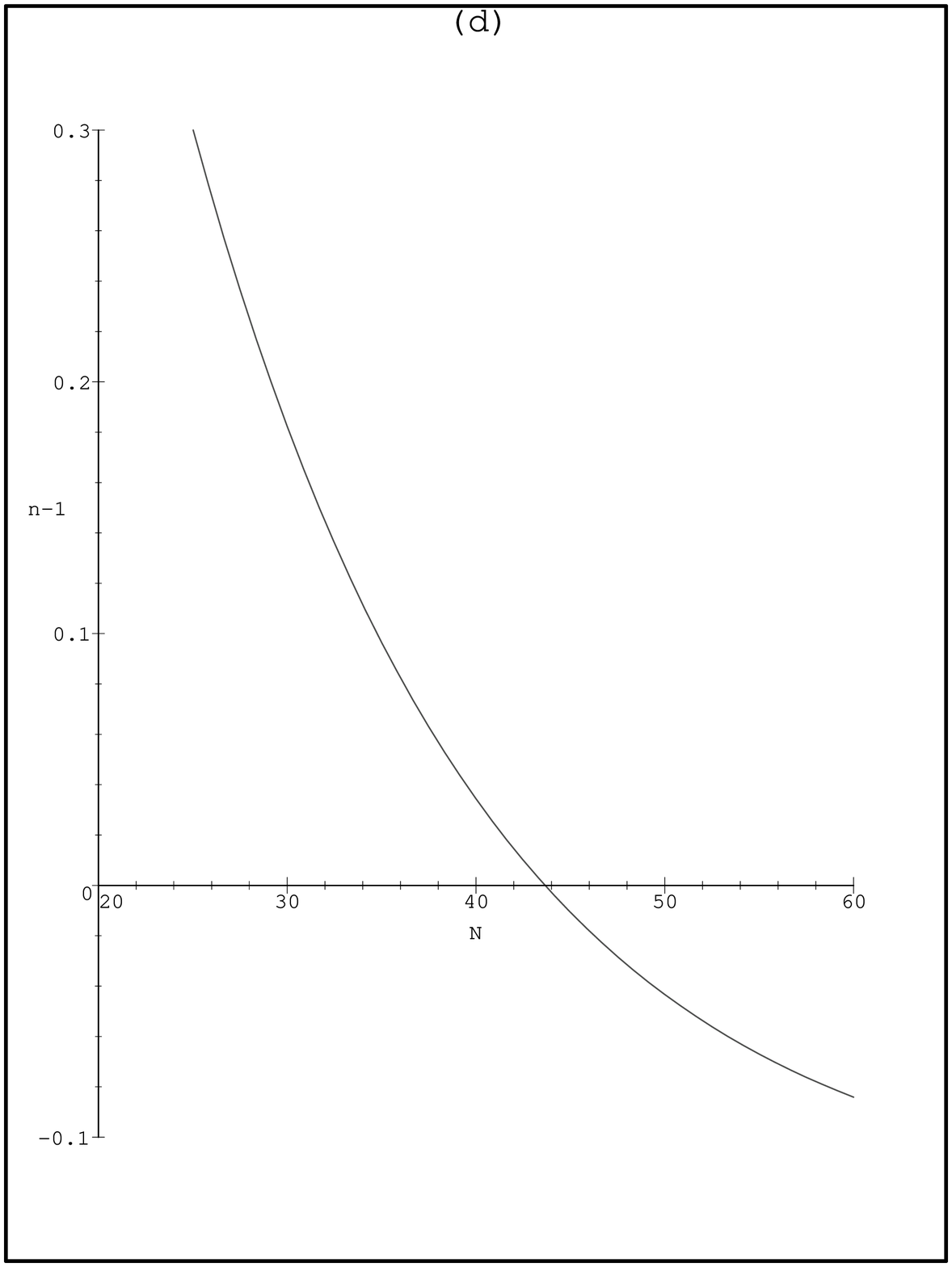}}
\caption{The spectral index as a function of comoving wavenumber.
The variable used is $N(\phi)\equiv \ln(k\sub{end}/k)$,
as described at the end of Section 4.
Cosmological scales are assumed to correspond to $40\lsim N(\phi)
\lsim 50$. 
Each plot corresponds to a representative point on the corresponding
plot of Figure 3; 
(a) $\tilde\alpha_0 = 0.1, A_0 = 0.2, \mu^2_0 = 0.1$; 
(b) $\tilde\alpha_0 = 0.05, A_0 = 0.4, \mu^2_0 = 0.2$; 
(c) for $\tilde\alpha_0 = 0.01, A_0= 1.7, \mu^2_0 = 1$;
(d) for $\tilde\alpha_0 = 0.001, A_0= 30, \mu^2_0 = 1.5$.}
\end{figure}

\section{Consistency constraints}

In addition to the observational constraints, we need to impose 
consistency conditions on the calculation. 

Two of these are the constraints already imposed, namely the condition 
$y_+ >y\sub{end}$ which corresponds to the statement that slow-roll 
inflation ends before the minimum represented by $y_+$ is reached,
and the condition that $y_\star $ is real representing the condition that
the potential possesses a maximum.

Since $\eta (y)$ is an increasing function of $y$, the first of these 
constraints is equivalent to
\be
1 = \eta (y\sub{end}) \leq \eta (y_+) = 3 A_0 \tilde\alpha_0 y_+^3 \left[ 
\tilde\alpha_0 y_+ - {2\over 3} \right].
\ee
Using expression~(\ref{roots}) for small $\tilde\alpha_0 y_\star $ and
approximating $y^2_\star$ with $y^2_{\star\star}$ keeping only the
leading order in the gauge coupling, gives the lower bound for $A_0$:
\be
A_0 \geq {\tilde \alpha_0^2 (1 + {21\over 4} \mu_0^2 ) \over 1-{21\over 4} 
\tilde\alpha_0^2} > \tilde \alpha_0^2 .
\label{bound-A-2}
\ee

The second requirement is equivalent to the statement that
\eq{gamma-y} has three real roots, which is
\be
\mu_0^2 \leq \left( {4\over 27 \tilde\alpha^2_0} -1\right) A_0.
\label{secondbound}
\ee

These two requirements are satisfied everywhere in the
regions of parameter space shown in Figure 3, except 
for Figure 3a where the first requirement corresponds to the nearly 
vertical line on the left hand side. They are plotted in Figure
5.

There are additional requirements, which also turn out not to be 
significant in practice. Let us mention them briefly.

Since we are using the one-loop approximation, we should require
$\tilde \alpha(Q)\ll 1$
for all $Q$ in the slow-roll regime $Q(=\phi) > \phi\sub{end}$.
But by virtue of the first line of \eq{roots}, this is ensured by
the condition $y\sub{end}<y_+$ that we imposed earlier.

\begin{figure}
\epsfysize= 2.2in %2.75in
\leftline{\epsfbox{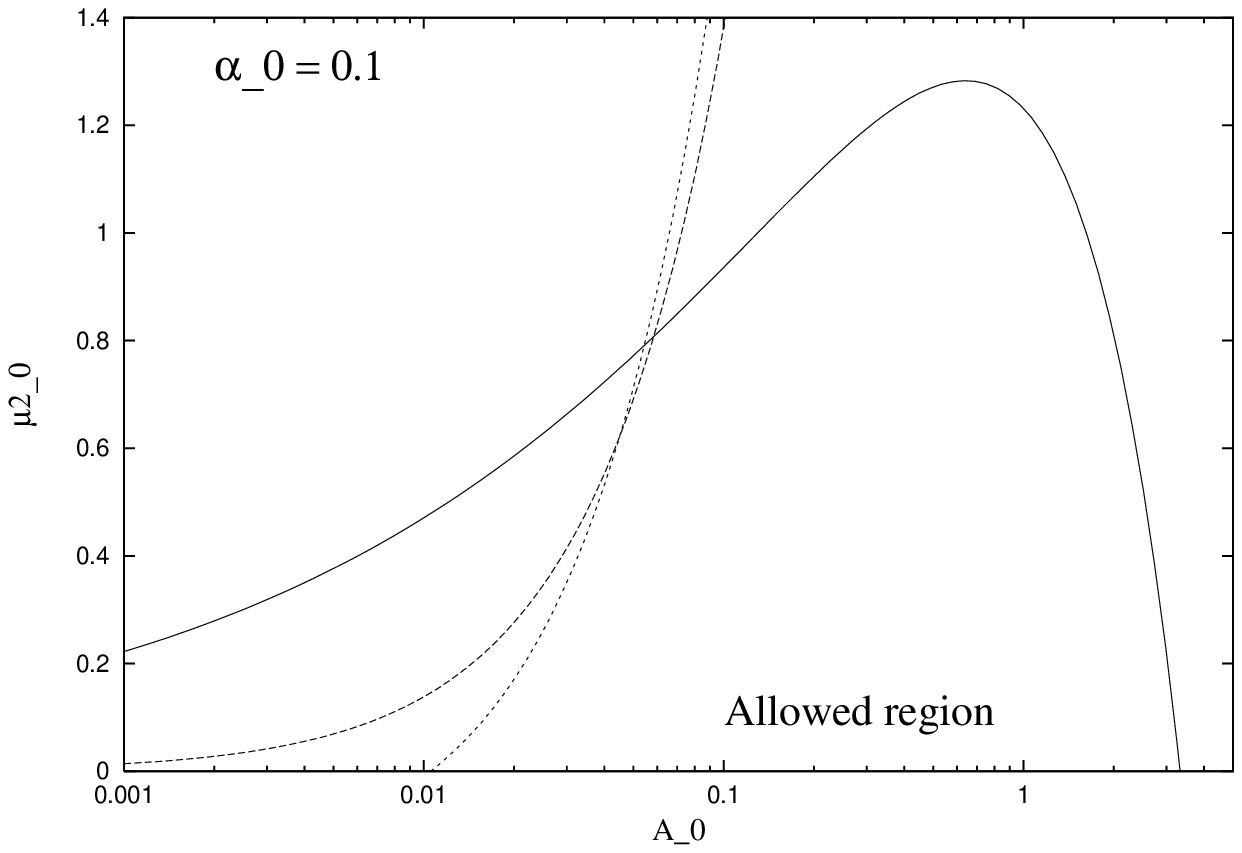}\epsfysize= 2.2in\epsfbox{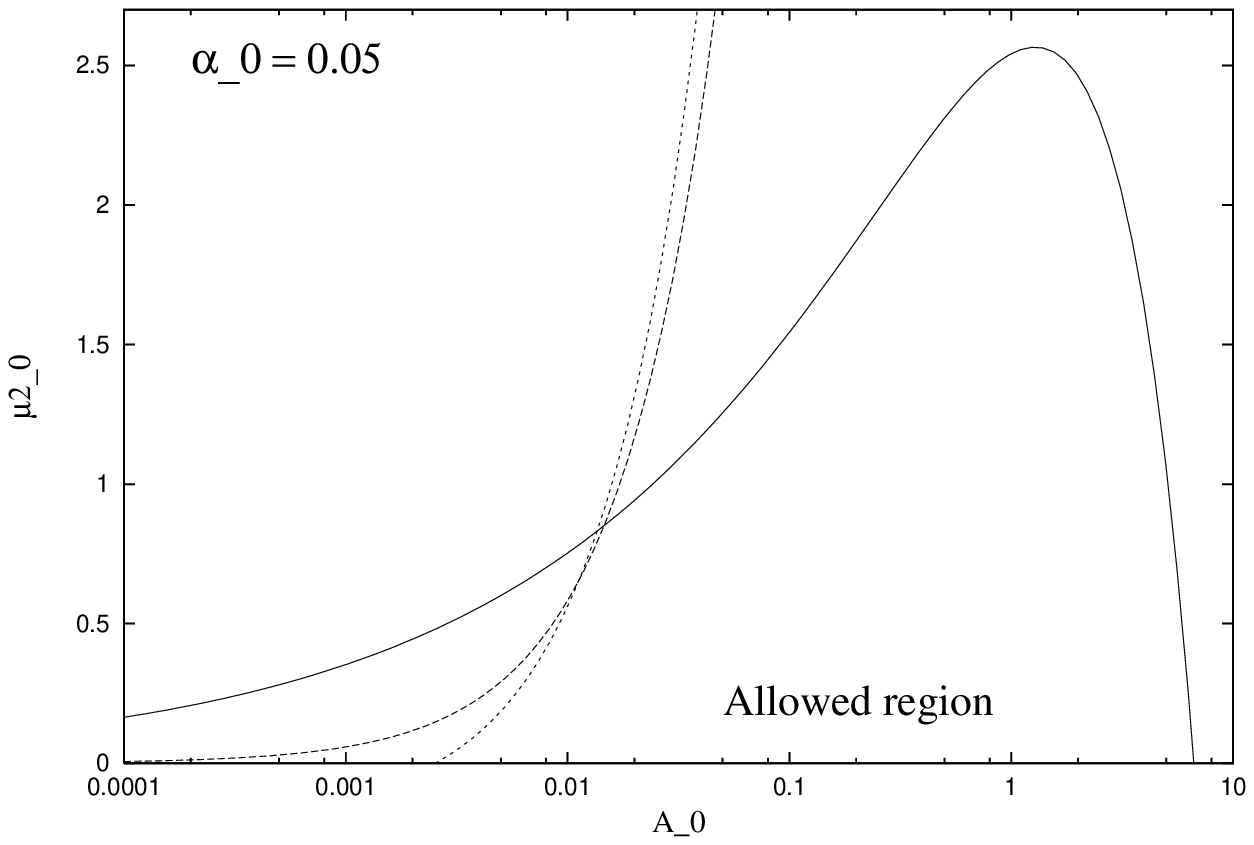}}
\epsfysize= 2.2in
\leftline{\epsfbox{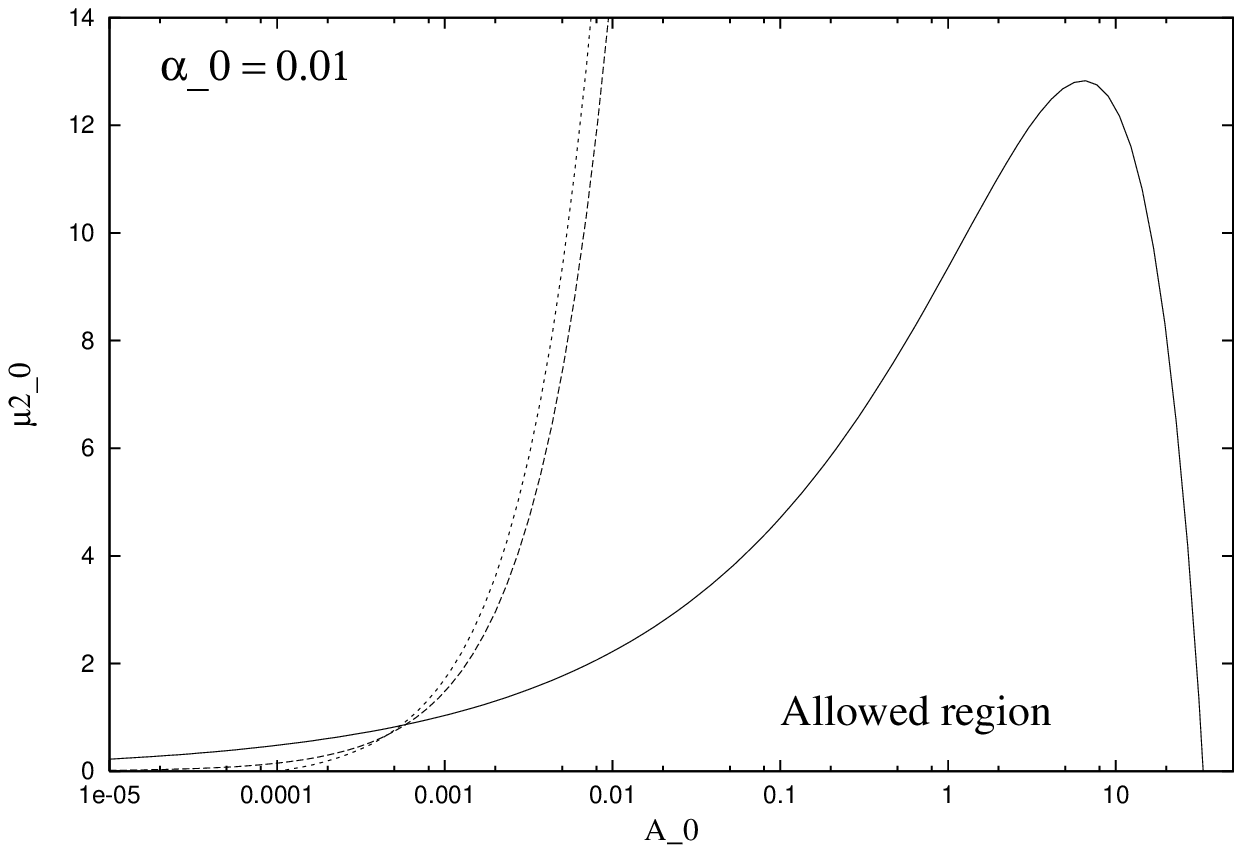}\epsfysize= 2.2in\epsfbox{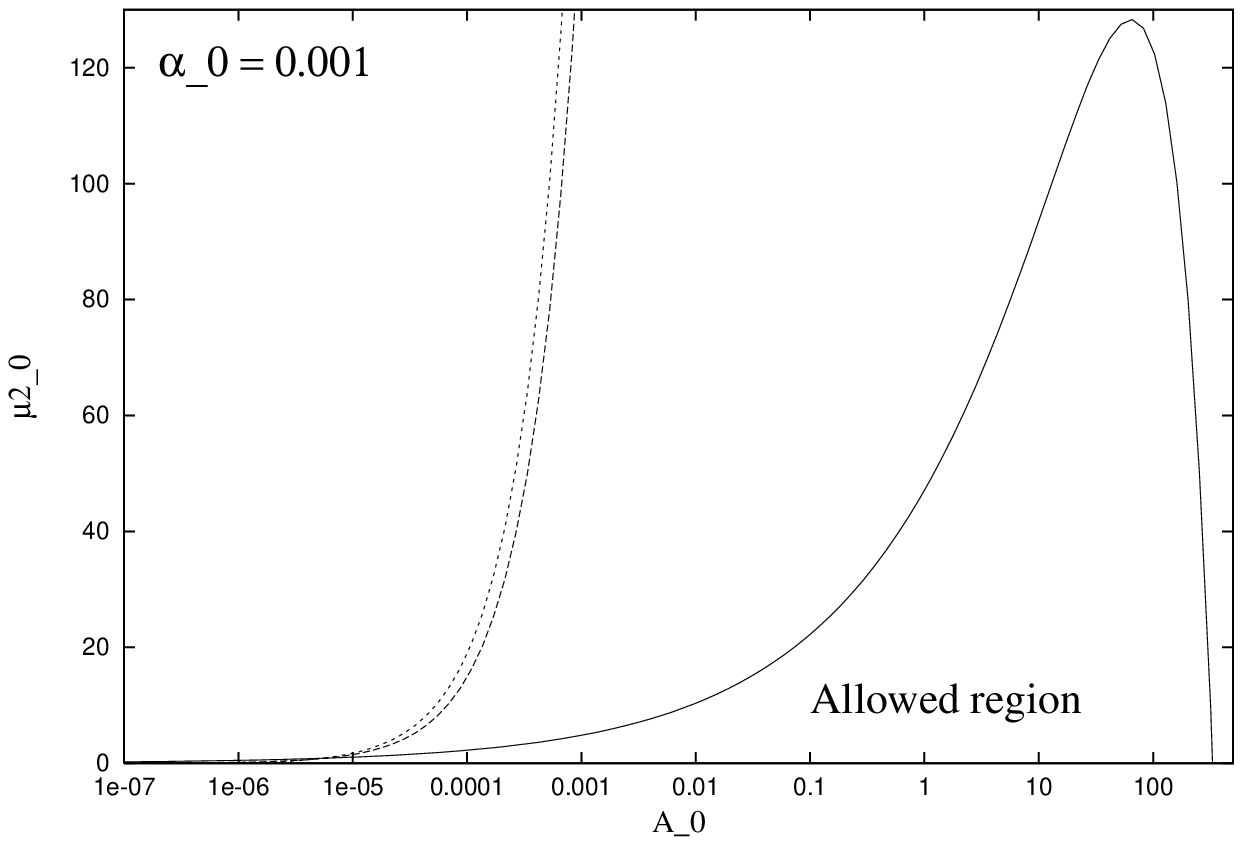}}
\caption{The region in the $\mu^2_0-A_0$ plane allowed by the
consistency checks for the different values of the coupling 
($\tilde\alpha_0 = 0.1,0.05,0.01,0.001$) is below the curves 
down to the x-axis. 
They represent Eqs.~(\ref{bound-A-2})
 (dotted line), (\ref{secondbound}) (dashed line)
and (\ref{bound-A-1}) (full line). }
\end{figure}

Another condition is the requirement $|\eta(\phi)|< 1$, which must hold
throughout inflation.
Recall that $\eta$ is a decreasing function of $\phi$, starting
at $+1$ when $\phi=\phi\sub{end}$, and ending at $\phi=\mpl$ with
some value $\eta(\mpl)\simeq \mu_0^2$. We are taking the latter value 
to be roughly of order 1, but if it bigger then the requirement
that $|\eta|<1$ throughout slow-roll inflation 
could in principle be violated. To be generous, let us impose this
requirement for all $\phi<\phi_{\star \star }$, 
which includes all $\phi<\phi_\star $ 
(the position of the maximum) and therefore all values of $\phi$ that we
could possibly be interested in.\footnote
{If $\phi$ arrives on the inflationary trajectory
by tunneling, at some value $\phi\sub{tunnel}$, we are interested
in $\phi<\phi\sub{tunnel}$. More generally \cite{p97toni}, 
note that slow-roll 
requires $\calp_\calr(k)\lsim 1$. This condition is violated above
some value $\phi\sub{eternal}$, and for bigger values
the random quantum motion dominates slow-roll, corresponding to what is 
called eternal inflation. If the inflaton starts out in the regime of 
eternal inflation, one is interested in $\phi<\phi\sub{eternal}$.}
This gives the relation
\be
\mu_0^2 \leq \left( {A_0\over 9 \tilde\alpha^2_0}\right)^{1/3} - A_0
\,,
\label{bound-A-1}
\ee
which is also shown in Figure 5.

A more involved requirement is
$\tilde m, m \ll g \phi $ for all values of $\phi$ during
slow-roll inflation, corresponding to \eq{tilmmreq}.\footnote
{In view of footnote \ref{gfoot} the factor $g$ is superfluous but we 
include it anyway.}
The most stringent bound is given by the 
inequalities evaluated at $y\sub{end}$:
\be
\tilde m^2_0 \ y^2\sub{end} = {|b|\over 2 c} V_0 A_0\ y^2\sub{end} \ll 
\alpha (y\sub{end}) \phi^2\sub{end} 
= {2\pi\over |b|} \tilde \alpha_0 y\sub{end} \exp\left[ -2
{ y\sub{end} - 1\over \tilde\alpha_0 y\sub{end} } \right] 
\label{bound-m}
\ee
and
\be
V_0 | \mu^2 (y\sub{end}) | =  V_0 A_0 (y\sub{end}^2 - y^2_{\star\star}) \ll 
\alpha (y\sub{end}) \phi^2\sub{end} = {2\pi\over |b|} \tilde\alpha_0 
y\sub{end} \exp\left[ -2 { y\sub{end} - 1\over \tilde\alpha_0 y\sub{end} } 
\right].
\label{bound-mu}
\ee
Assuming the COBE normalization, and $|b|$ and $c$ very roughly of order 1,
these bounds are weaker than the condition in eq.~(\ref{bound-A-1}) and 
therefore they are always satisfied in the region shown in Fig.~3a-d in the 
plane $\mu^2_0$ vs $A_0$ for different values of $ \tilde\alpha_0 $.
Notice that the larger is $\tilde\alpha_0$, the smaller the allowed 
region since most of the bounds scale as negative powers of the gauge 
coupling.

Finally, we consider the requirement that the quantum fluctuation of
the inflaton field does not come to dominate the classical slow-roll
behaviour before slow-roll ends. 
This is equivalent to a constraint on the magnitude of
the curvature perturbation calculated from slow-roll inflation,
$\calp_\calr(k)\lsim 1$.
We find that for typical values of the parameters,
this condition is satisfied until just before
$\phi$ reaches the value $\phi\sub{end}$, that corresponds to $\eta=1$.
Probably, this means that the quantum fluctuation never spoils slow-roll
inflation. On the other hand, such a big fluctuation might produce
dangerous black holes, so one probably ought to demand a tighter
upper limit than just $\calp_\calr(k)\lsim 1$. An investigation of 
this, and other aspects 
of the astrophysics, will be reported in a future publication.

\section{Discussion}

We have explored the parameter space of a model of inflation with 
a running inflaton mass, and the essential result is displayed in 
Figure 3. Up to group-theoretic factors that are expected to be of 
order 1, $\mu_0^2$ is the expected magnitude of the inflaton 
mass-squared at the
Planck scale, and $A_0$ is the expected value of the gaugino 
mass-squared, 
both in units of $V_0/\mpl^2$. With the assumptions mentioned in
Section 2, the expected magnitudes are $\mu_0^2\sim 1 $ and
$A_0\sim 0$ to $1$. Taking for definiteness an uncertainty $10^\Delta
=100$, the ranges 
$1\lsim \mu_0^2\lsim 100$ and $ 0\lsim A_0\lsim 100$
are regarded as reasonable. The range $\mu_0^2\ll 1$ is regarded as
{\em un}reasonable, and the purpose of the model is to avoid that range.

In Figure 3 we have plotted contour lines of the spectral index $n$,
and also of the potential $V_0$ in units of $\mpl^4$.
In both cases the COBE normalization has been imposed, at an epoch
$N=45$ $e$-folds before the end of inflation.

The
range allowed by observation is roughly the one between the contours
$n=0.8$ and $n=1.2$. (As we see in a moment, $n$ has considerable 
scale-dependence, so the observational bound $|n-1|<0.2$ should not
be taken too literally.)
Regarding the magnitude of the potential,
a fiducial value is 
$V_0^{1/4}\sim 10^{10}\GeV$, 
or $V_0/\mpl^4\sim 10^{-34}$, which 
as mentioned earlier corresponds to the assumption that supersymmetry
breaking in the vacuum is gravity-mediated, and of the same strength as 
it is during inflation. Replacing `gravity-mediated' by `gauge-mediated'
in the previous phrase, we can have 
$10^5\GeV \lsim
V_0^{1/4}\lsim 10^{10}\GeV$,
or $10^{-54}\lsim V_0/\mpl^4\lsim 10^{-34}$. If we allow the strength of 
supersymmetry breaking to be much less we can have smaller values
of $V$,
but presumably not by too many orders of magnitude.

The remarkable thing about Figure 3 is that the predictions are 
insensitive to $V_0$, if it is regarded as a free parameter 
in something like the above range. Couplings $\tilde\alpha_0\gsim
0.05$ are disfavoured because $\mu_0^2$ is much less than 1, and this 
conclusion holds even if $V_0$ is allowed to be arbitrarily large.
Couplings $\tilde \alpha_0\sim 10^{-2}$ to $10^{-3}$ are 
allowed, for any reasonable value of $V_0$.
The gaugino mass never becomes very small in the allowed region of parameter 
space, even thought that case might be allowed theoretically.
Finally, a very important conclusion is that {\em values 
$\mu_0^2\gsim 10$ are forbidden}, because they would require
unreasonably small values of $V_0$.

The last result means that the mechanism of a running inflaton mass
cannot be made extremely efficient. It works only if, in the absence 
of running, we would have $|\eta|\lsim 10$.
Recalling the discussion in Section 2, this means that mechanism works
only if there is no strong cancellation between the terms in \eq{vexp}.
Otherwise, the result $\mu_0^2\lsim 10$ requires 
\be
\frac{\msinf^4}{V_0} \lsim 10^{1+\Delta}\,.
\ee
where $10^\Delta$ is the uncertainty factor. It is clear that we can 
tolerate only a moderate degree of cancellation.

Subject to these constraints, this model of inflation
looks
quite attractive. With absolutely no fine tuning of parameters outside 
their expected  range, we can reproduce the COBE normalization and
keep the spectral index inside the observational bounds. Moreover,
the latter has significant variation on cosmological scales. 
Such a variation will be detectable in the forseeable future, and if
found it will strongly support the model.

\section*{Acknowledgements}
LR would like to thank KIAS, the Korea Institute for Advanced Study, for
kind hospitality and support where part of this work was done.

\def\NPB#1#2#3{Nucl. Phys. {\bf B#1}, #3 (19#2)}
\def\PLB#1#2#3{Phys. Lett. {\bf B#1}, #3 (19#2) }
\def\PLBold#1#2#3{Phys. Lett. {\bf#1B} (19#2) #3}
\def\PRD#1#2#3{Phys. Rev. {\bf D#1}, #3 (19#2) }
\def\PRL#1#2#3{Phys. Rev. Lett. {\bf#1} (19#2) #3}
\def\PRT#1#2#3{Phys. Rep. {\bf#1} (19#2) #3}
\def\ARAA#1#2#3{Ann. Rev. Astron. Astrophys. {\bf#1} (19#2) #3}
\def\ARNP#1#2#3{Ann. Rev. Nucl. Part. Sci. {\bf#1} (19#2) #3}
\def\mpl#1#2#3{Mod. Phys. Lett. {\bf #1} (19#2) #3}
\def\ZPC#1#2#3{Zeit. f\"ur Physik {\bf C#1} (19#2) #3}
\def\APJ#1#2#3{Ap. J. {\bf #1} (19#2) #3}
\def\AP#1#2#3{{Ann. Phys. } {\bf #1} (19#2) #3}
\def\RMP#1#2#3{{Rev. Mod. Phys. } {\bf #1} (19#2) #3}
\def\CMP#1#2#3{{Comm. Math. Phys. } {\bf #1} (19#2) #3}

\end{document}